\documentclass[11pt]{article}
\usepackage[a4paper,hmarginratio=1:1,vmarginratio=2:3,
totalwidth=14.7cm,totalheight=22.5cm]{geometry}
\usepackage{bm,epstopdf,epsfig,amsmath,amssymb,amsfonts,colordvi,latexsym,comment,cancel,
verbatim}
\usepackage{graphicx,graphics}
\usepackage[font=md,captionskip=8pt]{subfig}
\usepackage[usenames,dvipsnames]{color}

\usepackage{setspace}

\newcommand{\cD}{{\cal D}}

\newcommand{\cJ}{{\cal J}}

\newcommand{\cL}{{\cal L}}

\newcommand{\cO}{{\cal O}}

\newcommand{\Tr}{\mbox{Tr}}

\newcommand{\be}{\begin{equation}}
\newcommand{\ee}{\end{equation}}
\newcommand{\bea}{\begin{eqnarray}}
\newcommand{\eea}{\end{eqnarray}}
\newcommand{\Ra}{\Rightarrow}

\newcommand{\baa}{\begin{array}}
\newcommand{\eaa}{\end{array}}
\long\def\symbolfootnote[#1]#2{\begingroup
\def\thefootnote{\fnsymbol{footnote}}\footnote[#1]{#2}\endgroup}
\begin{document}
\begin{flushright}
CERN-PH-TH/2013-271.\\
\end{flushright}

\thispagestyle{empty}

\vspace{3.7cm}

\begin{center}
{\Large {\bf SUSY naturalness without prejudice}} 
\\
\medskip
\vspace{1.cm}
\textbf{D.~M. Ghilencea}$^{\,a, b,\,}$\symbolfootnote[1]{
E-mail address:  dumitru.ghilencea@cern.ch}

\bigskip
\medskip

$^a$ {\small CERN Theory Division, CH-1211 Geneva 23, Switzerland}

$^b$ {\small Theoretical Physics Department, National Institute of Physics and}

{\small  Nuclear Engineering (IFIN-HH) Bucharest, MG-6 077125, 
Romania.}

\end{center}

\bigskip
\def\baselinestretch{1.14}
\begin{abstract}
\noindent
Unlike the Standard Model (SM), supersymmetric models  stabilize the electroweak (EW) scale $v$ 
at the quantum level and {\it predict} that  $v$ is a function of the TeV-valued
SUSY  parameters  ($\gamma_\alpha$) of the UV Lagrangian. We show that the (inverse of the) 
covariance matrix of the model 
in the basis of these parameters and the usual deviation $\delta\chi^2$  (from $\chi^2_{min}$
of a model)  automatically encode information about the ``traditional'' EW  fine-tuning
measuring this stability, {\it provided that}  the EW scale $v\sim m_Z$ is indeed
regarded as a function $v=v(\gamma)$.
It is known that  large  EW fine-tuning may signal
 an incomplete theory of soft terms and can be reduced when 
relations among $\gamma_\alpha$ exist (due to GUT symmetries, etc).
The global correlation coefficient of this  matrix
can help one  investigate if such relations are present.
An upper bound on the usual EW fine-tuning measure (``in quadrature'') emerges from the 
analysis of the  $\delta\chi^2$ and the s-standard deviation confidence interval
by using $v=v(\gamma)$ and the theoretical approximation (loop order) considered
for the calculation of the observables. This upper bound avoids 
subjective  criteria for the ``acceptable'' level of  EW fine-tuning for which 
the model is still ``natural''.
\end{abstract}

\newpage

\section{A natural test for SUSY models.}

\subsection{Introduction}\label{s1}
Unlike the Standard Model (SM), supersymmetric models  (MSSM, NMSSM, etc)
stabilize the electroweak (EW) scale $v$  at the quantum level and 
make a  {\it prediction} for it: the combined higgses EW vev $v$ or
the Z boson mass  $m_Z\propto v$, is a  {\it derived} quantity that is a
function $v=v(\gamma)$ where $\gamma_\alpha$ denote
  the  Lagrangian UV parameters:
 the TeV-valued soft masses, soft couplings and $\mu$.
The function $v=v(\gamma)$ is obtained from the  minimization of the Higgs potential
and from the fact that the higgs couplings are fixed, to lowest order,
by gauge interactions (unlike the SM case where the higgs self-coupling is arbitrary).
Whether this prediction  successfully recovers the 
measured value  $m_Z^0\approx  91.187$ GeV
of the  $Z$ boson is  a {\it natural} test of SUSY.
This regards $m_Z$ as an {\it observable} to be fitted.
This view,  adopted here, remains true to the original motivation 
of  SUSY.

This naturalness test received much attention from  theorists
who long ago introduced fine-tuning \cite{susskind} measures 
\cite{finetuning,castano} for it. However, precision data fits 
\cite{Bechtle}
often prefer  to  keep  $m_Z$  as a {\it fixed input} (constant) equal to $m_Z^0$
rather than as an observable  as well that depends on the Lagrangian parameters
and then no  $\chi^2$ ``cost'' to fit $m_Z$ is 
usually reported\footnote{This is partly due to technical and
historical reasons from pre-SUSY fits where $m_Z$ is an input; with $m_Z$ output
also the parameter scans would be very ineffective, as many  points are ruled
out by $Z$ mass.}.
At the same time, some  data fits still report the EW fine-tuning 
in which  $v\sim m_Z$ is indeed a function $v=v(\gamma)$, often giving
 a large  variation of $m_Z$ (about its fixed input value). 
It is not clear to us how
 results relying on two different assumptions ($v$  {\it fixed} constant
or a function $v=v(\gamma)$)  can be combined consistently to draw a 
clear conclusion.  This is due to the following questions 
(particularly $Q_3$ below):

\noindent
$Q_1$: Is  the likelihood to fit the observable $m_Z$ 
related to the EW fine-tuning ``cost''? 

\noindent
$Q_2$: What is the link of the 
 total likelihood  to fit a set of observables 
to  EW fine tuning?

\noindent
$Q_3$: How do we compare a model with a good fit (of $m_Z$ and other data)
and ``large'' EW fine tuning to one with
nearly-as-good a fit but less fine tuning (for the same data)?

Question $Q_1$  is  even more compelling given that we do not know what
an  ``acceptable'' value of the EW fine-tuning is.
One can  address $Q_{1,2,3}$  by regarding $m_Z$ as an observable and by 
using standard tools to test models, as discussed in the likelihood approach
 \cite{Ghilencea:2013hpa,Ghilencea:2013fka}
or earlier in the Bayesian case \cite{Casas,Fichet:2012sn,Ghilencea:2012gz}.
These works  suggested that  EW fine-tuning is
related to  the likelihood ($\chi^2$) or the posterior 
probability to fit the data that  includes the  observable  $m_Z$. 
In this letter we  explore this relation further, using a different approach,
and a $\chi^2$ (frequentist) analysis.
As detailed below, we study a possible  relation of the  
covariance matrix of the model to the EW fine-tuning. 
This connection is not examined in the literature even though each of these aspects
were studied in the past separately.
This is the main purpose  of this work.

As an example, consider the  MSSM case  
with the higgs potential minimum conditions in a standard notation, 
fixing the EW scale ($v\propto m_Z$) and $\tan\beta$ (or $B$):
\bea\label{mm}
\frac{m_Z^2}{2}=-\mu^2+\frac{m_1^2-m_2^2\,\tan^2\beta}{\tan^2\beta-1}
+\dots
\nonumber\\[2pt]
2\, m_3^2=(m_1^2+m_2^2+2\mu^2)\,\sin 2\beta
+\cdots
\eea
The dots stand for quantum corrections to the quartic higgs couplings
and $m_{1,2,3}$ are one-loop soft masses.
In precision data fits, 
one traditionally replaces $m_Z$ ``by hand'' by its  measured mass $m_Z^0$ 
to fix  $\mu$ instead (as a function of remaining $\gamma_\alpha$),
or ``fine-tune''  the independent $\gamma_\alpha$
to reproduce  $m_Z^0$. Ultimately, this  amounts to using a 
Dirac $\delta$-distribution for the observable  $m_Z$.
Further \cite{Ghilencea:2013hpa,Ghilencea:2013fka}
\bea
\qquad
\delta(1-m_Z/m_Z^0)= 
\frac{1}{\Delta}\,\delta\big[n^\alpha (1-\gamma_\alpha/\gamma_\alpha^0)\big], 
\quad 
\gamma=\{m_0, \mu, A_0, B_0, M_1, M_2, M_3,...\}
\label{zm}
\eea
Here $m_Z\sim v$ is a function of parameters $\gamma_\alpha$ as shown in  eq.(\ref{mm}),
with $\gamma_\alpha^0$ components of the set $\gamma$
that respect the condition  $m_Z(\gamma_\alpha^0)=m_Z^0$  and
$n^\alpha$ are the components of the normal to the surface defined by this equation; finally
\bea\label{Delta}
\Delta\equiv
\Big\{\sum_\alpha \Big(\frac{\partial 
\ln m_Z(\gamma)
}{\partial \ln \gamma_\alpha}
\Big)^2_{\gamma=\gamma^0}\Big\}^{1/2}
\eea
Since $\Delta$ emerged from  fixing the EW scale condition ($m_Z=m_Z^0$)
associated with fine-tuning,
we can only  interpret it as a {\it derived}, unique measure of fine-tuning 
({\it not} chosen)!

The  message is that the   distribution in the lhs of eq.(\ref{zm})
 chosen as a  likelihood (for observable $m_Z$) 
somehow ``knows'' about the EW fine-tuning $\Delta$.
This may not be too surprising, but it  hints to a deeper connection. 
The EW scale $v=v(\gamma)$ enters in many  observables and also correlations
among these can be present. Therefore they can also 
have  significant individual fine-tunings associated. 
We usually refer to fine-tuning of $m_Z$, but one can similarly
discuss, for example, the fine-tuning of the Higgs mass, etc, since this is
 also  closely related
to the hierarchy problem! Then a relation similar to eq.(\ref{zm}) 
can be present between each observable and
the set of parameters $\gamma$.
This seems to suggest an underlying connection of the
 likelihood (or usual $\chi^2$) associated with a set of EW observables to 
their fine-tuning  and  to the  distribution
of the parameters of the model about their central values (of maximal likelihood
or min $\chi^2$); 
these could be connected as in the example
above, by some general form of $\Delta$. 
These comments indicate an affirmative answer to question $Q_{1}$.
For a more detailed discussion see \cite{Ghilencea:2013hpa,Ghilencea:2013fka}.

In the following we explore some of these issues further. We consider that:

{\bf 1)} in practice
one does not have Dirac  distributions for $m_Z$ or other  observables; 

{\bf 2)}  correlations can exist  between $m_Z$ and other 
observables: $m_h$, $m_H$, etc. 

{\bf 3)} other observables can also depend on the EW
scale $v=v(\gamma)$ (and/or on $\gamma$).

\noindent
With these in mind  we study the link of the likelihood to fit the data and its
deviation from the maximal value, to the EW fine-tuning ($\Delta$ above);
 equivalently, in a chi-square language, we study the link of the deviation
$\delta\chi^2$ from the minimal value $\chi^2_{min}$, to the EW fine-tuning, $\Delta$.
We find that, just as the likelihood to fit $m_Z^0$ 
contains $\Delta$  (eq.(\ref{zm})), 
in the general case (with {\bf 1)}, {\bf 2)}, {\bf 3)}) of the total likelihood,
EW fine tuning is automatically present in 
the  covariance matrix $\tilde M$ 
in the basis of the fundamental parameters  $(\gamma_\alpha)$ and thus also
in the deviation $\delta\chi^2$, 
{\it provided that} we regard $v$ as a function $v\!=\!v(\gamma)$,
(as predicted by SUSY).
Thus the matrix $\tilde M$  has a more fundamental role than the EW 
fine-tuning and contains information about the stability of the EW scale under
UV variation of SUSY parameters.
This view also  ends a long-held distinction between 
EW fine tuning (to fit $m_Z$) and that  to fit other 
 observables ($m_h, m_H$, etc) that also  depend  on $v$  
and  that are thus ultimately linked to the hierarchy problem (just as $m_{Z, W}$
are).

\subsection{The link of  the likelihood ($\chi^2$)  to  EW fine-tuning.}
\label{qqq}

Consider a  model with a number of observables $\cO_i$ ($i=1,2,...,n$) of central
experimental values $\cO_i^0$ fitted using a 
set of SUSY parameters\footnote{$\gamma$ can 
 include nuisance variables (Yukawa couplings, etc)
eliminated later by integration/profiling.} 
$\gamma_\alpha$, ($\alpha=1,2.., s$)  that enter in the
Lagrangian with $s<n$ and $n_{df}=n-s$ ($n_{df}$: number of degrees of freedom).
The general form of  eq.(\ref{mm}) of the two minimum conditions of the scalar potential is
\medskip
\bea\label{gs}
 v=v(\gamma;\beta),\qquad 
\tan\beta=\tan\beta(\gamma, v) 
\qquad
\gamma_\alpha: m_0, \mu, A_0, B_0, M_1, M_2, M_3,....
\eea

\medskip\noindent
leading to\footnote{One often replaces $\beta$ by some other parameter, like
$B_0$, with no implications below.}  $v=v(\gamma, \beta(\gamma))$; to  simplify the notation, 
hereafter we refer to this dependence  as $v(\gamma)$.
From a Taylor series for  $m_Z\!\sim\! v$ about a particular point $\gamma^0$:
\medskip
\bea\label{ttt}
m_Z=m_Z(\gamma^0)+\Big(\frac{\partial m_Z}{\partial \gamma_\alpha}\Big)_{\gamma=\gamma^0}
\,(\gamma_\alpha-\gamma_\alpha^0)+\cdots,
\eea

\medskip\noindent
 Assume for a moment  that $\gamma^0$ is a 
solution\footnote{Note that with an
over-constrained set of parameters $\gamma_\alpha$ by the
set of observables ($n_{df}>0$), $\gamma^0$ above should actually
denote the set that minimizes the global $\chi^2$ of all observables, 
including  $m_Z$, rather than the solution to $m_Z(\gamma)=m_Z^0$
(see Section~\ref{ffrr}).
Then $m_Z(\gamma^0)$ does not reproduce 
the central measured value, but a value that should be within
few standard deviations from it (say $2\sigma_z$); then
 the difference in the lhs of  (\ref{r1}) should again be understood
 as $2\sigma_z$.}
to $m_Z(\gamma)=m_Z^0$, with $m_Z^0\approx  91.187$ GeV.
Eq.(\ref{ttt}) can be re-written as
\medskip
\bea\label{r1}
\frac{m_Z-m_Z(\gamma^0)}{m_Z(\gamma^0)}=\Delta\, n^\alpha\,
\frac{\gamma_\alpha-\gamma_\alpha^0}{\gamma_\alpha^0}+\cO((\gamma_\alpha-\gamma_\alpha^0)^2),
\qquad
n^\alpha\equiv\frac{1}{\Delta}
\Big(\frac{\partial \ln m_Z}{\partial\ln \gamma_\alpha}\Big)_{\gamma=\gamma^0}; 
\eea

\medskip\noindent
where $n^\alpha$ denote the components of the 
normal to the surface $m_Z(\gamma^0)=m_Z^0$, with $n_\alpha n^\alpha=1$.
Here $\Delta$ is given in eq.(\ref{Delta}) with eq.(\ref{gs}).

For the recently measured $m_h\approx 126$ GeV \cite{SMH} we know that
  minimal $\Delta$, upon
 varying all allowed parameters $\gamma_\alpha$ and $\tan\beta$,
is $\Delta\sim \cO(1000)$  \cite{Ghilencea:2012gz} in the  MSSM-like models
with different boundary conditions for the soft terms. Note however that
in the general version of the NMSSM (GNMSSM) a value of $\cO(20)$ is still possible
 \cite{GGR}.
Then to keep $m_Z$ of eq.(\ref{r1}) within $2\sigma_z$ 
of its central measured value\footnote{as usually done in the data fits  for any observable.}
$m_Z^0$  or equivalently $\delta m_Z/m_Z^0=4.6\times 
10^{-5}$, one must keep each parameter $\gamma_\alpha$ within an order of
$\delta\gamma_\alpha/\gamma_\alpha^0\approx 4.6\times 10^{-8}$ of its value $\gamma_\alpha^0$. 
For simplicity, assume that
 all parameters other than one of them, say $\mu$, are fixed and take for example
 $\mu_0\approx 1$ TeV, so it would mean $\delta\mu=46$ keV.
Such accuracy $\delta\mu$ or $\delta\gamma_\alpha$ needed
  to compensate the large $\Delta$ in the rhs of (\ref{r1}), 
 can be reached by a  fine scan of the parameter 
space; but  a deviation by few keV 
 deviates $m_Z$ by more than $2\sigma_z$ from its measured value, 
if  $\Delta\approx 1000$. 
So a good stability of the $\chi^2$ fit of $m_Z$  and a large $\Delta$,
are not  easily compatible 
(also recall that $\gamma_\alpha^0$
are (over)constrained to keep under control $\chi^2$ due to  other observables).
This  problem reflects a relation of the  ``traditional'' fine-tuning of eq.(\ref{Delta})
and the associated $\chi^2$ ``cost'' to fit $m_Z$ or other observables
that depend on $v(\gamma)$. If one insists
of keeping $m_Z$  a fixed input number (equal to  $m_Z^0$), 
the problem remains because in the above
 discussion one replaces $m_Z$ by any another observable that depends on $v=v(\gamma)$.
Therefore, the relation of $\Delta$ to $\chi^2$ is important and usually
overlooked in the literature.

If relations among initial $\gamma_\alpha$ exist, dictated for example by UV
symmetries (SU(5), etc), they can reduce\footnote{$\Delta$ can also be reduced
by ``new physics'' in the Higgs sector that can increase the  Higgs  mass
\cite{GGR,extmssm}.} $\Delta$ \cite{Ghilencea:2012gz,su5}. 
So a  large $\Delta$ can simply be a sign of our ignorance of 
the UV physics, telling us that our theory of soft terms is  inappropriate
or incomplete.
Aside from this possibility, dramatic fine-tuning of $\gamma_\alpha$ could be ``natural'' 
if $\gamma_\alpha$ are related to a {\it fundamental} constant of Nature,
 whose accurate determination is crucial for the theory.

There are however limits to how much one can ``fine-tune'' 
$\gamma_\alpha$ in a given loop order.
Indeed, $\gamma^0_\alpha$ are  determined 
from  the condition of minimizing  total $\chi^2$ computed 
using a theoretical calculation of the
observables in  a fixed loop-order. This calculation is affected 
by an error from ignored  higher loops 
(an example  is the 2-3 GeV theoretical error of the Higgs mass
 at 2-loop \cite{theor-higgs-err}). 
So the perturbation theory alone inevitably introduces a 
theoretical error $\sigma_{th}$ to each $\gamma_\alpha^0$.  
Then only points with $\sigma_{\gamma_\alpha}>\sigma_{th}$ are actually relevant\footnote{
A naive estimate of $\sigma_{th}$ can be, at 3 loop order, $1/(16\pi^2)^3$, 
which is larger than $4.6\times 10^{-8}$ mentioned. 
More correctly, $\sigma_{th}$ of each parameter
is found numerically from the error of ignored loops in the theoretical value of the observables
(such as that of $m_h$ mentioned) that depend on that  parameter.};  with this bound,
from eq.(\ref{r1})  $\Delta$ then  has an upper bound if one insists to
keep $m_Z$ within  say $2\sigma_z$ from its central value.
So $\delta\chi^2$ and $\Delta$ are related.

The above discussion can be extended to all observables that depend on the 
EW scale $v(\gamma)$; then a relation 
between each observable and the amount of tuning of $\gamma_\alpha$
 is present, like in eq.(\ref{r1}), 
with a similar  connection to their  $\chi^2$ contribution\footnote{In 
this case $\gamma_\alpha^0$ will correspond to the  maximal likelihood point.}.
It is then natural to expect a more general connection between
the  total $\chi^2$ (or more generally, the  likelihood) of all observables
including $m_Z$  and  their fine-tuning with respect to  $\gamma_\alpha$. The 
generalisation of $\sigma_\gamma$ and $\sigma_z$ 
discussed above is  the covariance matrix, therefore the latter
could be the missing link in this connection (see later,
Section~\ref{ffrr}).


\subsection{Fixing the EW scale and the relation to $\Delta$.}

Let us denote by $L(\cO\vert \gamma)$  the total likelihood 
to fit some observables $\cO_i$ other than $m_Z$.
We impose on this likelihood the condition of fixing the EW scale 
that motivated SUSY,
that we  regard just as a condition to fit $m_Z\sim v$ of eq.(\ref{mm})
to its central measured value. We first take a Dirac delta 
distribution for $m_Z$. 
Assuming that we can factorize this distribution\footnote{We 
relax this assumption in Section~\ref{ffrr}.} 
from $L(\cO\vert \gamma)$ of other data,
  the total likelihood that accounts also for fixing the EW scale
is 
$L(\cO, m_Z^0\vert\gamma)=
L(\cO\vert \gamma)\,L(m_Z^0\vert\gamma)
$ 
with \cite{Ghilencea:2013hpa,Ghilencea:2013fka}
\medskip
\bea\label{LW}
L(m_Z^0\vert\gamma)=
\delta\Big(1-\frac{m_Z(\gamma)}{m_Z^0}\Big)
= 
\frac{1}{ \Delta}\,\delta 
\big( n^\alpha (1- \gamma_\alpha/\gamma_\alpha^0)\big)
\eea

\medskip\noindent
In the last step we used eq.(\ref{r1}), with $\Delta$ as in  eq.(\ref{Delta}).
The argument of the Dirac $\delta$ function  ensures that one must be on the surface
predicted in SUSY  by the minimum condition $m_Z(\gamma^0)=m_Z^0$ giving
 $n^\alpha (1- \gamma_\alpha/\gamma_\alpha^0)=0$, 
or $\gamma_\alpha=\gamma_\alpha^0$. For a detailed discussion and interpretation of 
 eq.(\ref{LW}) see \cite{Ghilencea:2013hpa,Ghilencea:2013fka}
\footnote{
As a side-remark, eq.(\ref{LW}) can be 
formally integrated in one general direction (combination of $\gamma_\alpha$)
and re-written as  \cite{Ghilencea:2013hpa}: 
$\tilde L(m_Z^0\vert\gamma)=(1/\Delta)\vert_{\gamma_\alpha=\gamma_\alpha^0}$.
So $1/\Delta$ that emerged is  the likelihood ``cost'' of  respecting the SUSY  
condition of fixing the EW scale  and is part
of  total $L(\cD, m_Z^0\vert\gamma)$ to fit the data 
that includes $m_Z^0$. A similar interpretation
was noticed on phenomenological grounds in \cite{strumia}.
With this,  one can provide a very simple  estimate
 of an upper bound on $\Delta$. The contribution
to total $\chi^2$ due to $m_z$ alone (say $\delta\chi^2$) equals, according
 to the last equation $\delta\chi^2=-2\ln \tilde L=-2\ln(1/\Delta)$ (under some assumptions).
By demanding a ``good fit'' i.e. that 
{\it total} $\chi^2/n_{df}\approx 1$, ($\chi^2$ includes contributions from other observables)
one has $\Delta<\exp(n_{df}/2)$ \cite{Ghilencea:2013hpa,Ghilencea:2013fka}, which with 
usual $n_{df}\sim 10$ gives an upper value $\Delta \sim 100$.
One objection to this approach is that  $\tilde L$ is not 
normal which affects  the goodness of the fit criterion
 $\chi^2/n_{df}\sim 1$.}\textsuperscript{,}\footnote{
  The minimal value of $\Delta$ (with all $\gamma_\alpha$
allowed to vary), grows 
with the higgs mass $\Delta\sim \exp(m_h/{\rm GeV})$,
see figures 1-8 and 13-16  in \cite{Ghilencea:2012gz}, also \cite{Cassel:2009cx}. As a result 
an error $\delta m_h =2-3$ GeV that is the theoretical uncertainty of $m_h$ prediction
 \cite{theor-higgs-err} brings an uncertainty factor
 $\approx \exp(3)\approx 20$ for $\Delta$ and accordingly for its $\delta\chi^2$ effect.
This uncertainty means that  $\Delta\approx 20$ and $\Delta\approx 400$ 
can be seen as equally `acceptable'.
Thus the results   \cite{Ghilencea:2013hpa,Ghilencea:2013fka}
should be regarded as a general estimate rather than a strict
criteria of viability.}.
Further, it is  illustrative to  go beyond the Dirac $\delta$ used
in eq.(\ref{LW}), so one can take 
\medskip
\bea
L(m_Z^0\vert\gamma)=\frac{m_Z^0}{\sqrt{2\pi}\,\sigma_z}
\exp\Big(-\frac{m_Z^{0\, 2}}{2 \sigma_z^2}\,(m_Z/m_Z^0-1)^2\Big)
\eea

\medskip\noindent
which recovers the lhs of eq.(\ref{LW})  when $\sigma_z\rightarrow 0$.
 From the Taylor expansion in eq.(\ref{ttt}) about  $\gamma_\alpha^0$ 
 one finds, to a first approximation, with $\Delta$  as in eq.(\ref{Delta})
\medskip
\bea\label{zz}
L(m_Z^0\vert\gamma)
=
\frac{1}{\Delta}\,L(\gamma)+\cdots \quad \Ra\quad
L(\cO, m_Z^0\vert\gamma)=\frac{1}{\Delta}\,L(\gamma)\,L(\cO\vert\gamma)+\cdots
\eea

\medskip\noindent
The first equation  is similar to eq.(\ref{LW}). $L(\gamma)$ is the associated
normalized distribution of the output values of 
parameters $\gamma$; if all $\gamma$ are fixed to 
$\gamma^0$ except one of them ($\gamma_\alpha$), then
\smallskip
\bea
L(\gamma)=\frac{1}{\sqrt{2\pi}\,\sigma_{\gamma_\alpha}}
\exp\Big(-\frac{1}{2\,\sigma_{\gamma_\alpha}^2}\,
\big(\gamma_\alpha/\gamma_\alpha^0-1\, \big)^2
\Big)
\eea
with estimated 
\bea\label{sigma}
\qquad\qquad
\sigma_{\gamma_\alpha}=\frac{1}{\Delta}\frac{\sigma_z}{m_Z^0} \,\gamma_\alpha^0;
\qquad 
{\rm if}\,\,\sigma_{\gamma_\alpha}\geq \sigma_{th}
\,\,
\Ra\,\,
\Delta\leq \frac{\sigma_z}{\sigma_{th}}\frac{\gamma_\alpha^0}{m_Z^0}.
\eea

\medskip\noindent
$\Delta$ that emerged in (\ref{zz}), (\ref{sigma})  is the sole consequence
of the condition of fixing the EW scale ($m_Z=m_Z^0$) that is usually
 associated with fine-tuning,
 so it is a unique, derived measure (of fine-tuning) from this constraint!
$\Delta$ also relates the normalized 
likelihood (of $m_Z$) to the normalised Gaussian distribution  of  
$\gamma_\alpha$, about   the central value $\gamma_\alpha^0$; such 
relation is  more generic (see later).
Since $\Delta$ enters in the expression of  $\sigma_{\gamma_\alpha}$, 
this suggests again that in more complex cases
 the generalization of $\sigma_{\gamma_\alpha}$, the error matrix, could 
be related to the EW fine-tuning.
Finally, if one demands $\sigma_{\gamma_\alpha}\geq \sigma_{th}$  (from the
loop-order accuracy $\sigma_{th}$ that affects  $\gamma_\alpha^0$)  
an upper bound on $\Delta$ emerges in eq.(\ref{sigma}).
This is actually a strong bound, even assuming $\sigma_{th}\sim \sigma_z$,
then  $\Delta\leq \cO(10)$ for TeV valued $\gamma$'s.

\subsection{The general case: more observables, correlations
 and fine-tuning.}\label{ffrr}

So far we ignored the  correlations  of $m_Z$ with other  
observables (for example with the loop-corrected higgs masses  $m_h$, $m_H$)  
or the fact that other observables also depend on the EW scale.
We include these effects to examine the relations of
the   total likelihood ($\chi^2$) of these observables,
  its deviation from its maximal (min $\chi^2$) value, and
 of the covariance matrix, to the  EW fine-tuning.
The study  is restricted to Gaussian distributions for observables
so is  equivalent to a simple $\chi^2$-analysis with $\chi^2\!=\!-2\ln L$;
 the analysis can be extended to  general likelihoods.

Consider the observables $\cO_j$, $(j=1,2....n)$, of 
experimental central values $\cO_j^0$,  and to simplify the
notation we now assume that $m_Z$ is also one of them,   $\cO_n\!=\!m_Z$; they
are functions of $\gamma_\alpha$, $(\alpha=1,2,...,s)$, so $\cO_i\!=\!\cO_i(\gamma,v(\gamma))$, 
with $v(\gamma)$ as in eq.(\ref{gs}).
The total likelihood $L(\cO \vert\gamma)$ due to all $\cO_i$
must then be maximized with respect to the SUSY  parameters 
$\gamma_\alpha$. 
It is convenient to work with a dimensionless form of
this likelihood\footnote{
The ``dimensionful'' form of the total likelihood is
$\cL(\cO\vert\gamma)=(2\pi)^{-n/2} (\det K)^{-1/2}
\exp\big[-{1}/{2}\,\,(\cO_i-\cO_i^0)\,(K^{-1})_{ij}\,(\cO_j-\cO_j^0)\big]
$;
$K$ is the  dimensionful covariance matrix: $K_{ij}=\sigma_i\,\sigma_j\,\rho_{ij}$;
$\rho_{ij}=\rho_{ji}$ account for correlations; $\rho_{ii}=1$.
$\cL$ is equal to  $L$ used in the text up to a constant:
 $L(\cO\vert\gamma)=\!\vert\cO_1^0\cO_2^0....\cO_n^0\vert\times
\cL(\cO\vert\gamma)$.
} 
\medskip
\bea\label{ttrr}
L(\cO\vert\gamma)=(2\pi)^{-\frac{n}{2}}\, (\det M)^{-\frac{1}{2}}
\,\exp\big[-1/2\,\,u_i\,(M^{-1})_{ij}\,u_j\big],\qquad
u_i\equiv\cO_i/\cO_i^0-1.
\eea

\medskip\noindent
with a (dimensionless) covariance matrix 
$M_{ij}\!
=\!\rho_{ij}\,\sigma_i\sigma_j/(\cO_i^0\cO_j^0)$, 
$\rho_{ij}\!=\!\rho_{ji}$ denote the correlation coefficients, with $\rho_{ii}\!=\!1$.
Let  $\gamma^0$ denote the solution of the condition to maximize  $L(\cO\vert\gamma)$.
Although not appropriate for high precision numerical studies, to illustrate the 
main idea a Taylor expansion can be used 
\medskip
\bea\label{ss}
\cO_i(\gamma)=\cO_i(\gamma^0)+(\gamma_\alpha-\gamma_\alpha^0)\,
\Big(\frac{d \cO_i}{
d\gamma_\alpha}\Big)_{
\gamma=\gamma^0}+\cdots,
\eea
then
\bea\label{log}
L(\cO \vert\gamma)
=\frac{\kappa}{\Delta}\,\,L(\gamma)+\cdots
\eea
where $\kappa$ is  a constant and\footnote{In a $\chi^2$ language:
$\kappa\!=\!(2\pi)^{-n_{df}/2}\exp(-\chi^2_{min}/2)$ with 
$\chi^2_{min}=\!u_i(\gamma^0)\, M^{-1}_{ij} \,u_j(\gamma^0)$ and
$\delta\chi^2\!=\!\chi^2\!-\!\chi^2_{min}$
with $\delta\chi^2\!=-2\ln \big[ L(\cO\vert\gamma)/L(\cO\vert\gamma^0)\big]=-2\ln
\big[L(\gamma)/L(\gamma^0)\big]$.} 
\bea\label{pp}
\Delta\!\equiv \!
\big[\det M \det \tilde M^{-1}\big]^{\frac{1}{2}}\!,
\quad
\tilde M^{-1} \equiv \cJ^T M^{-1}\cJ,
\quad
\cJ_{i\alpha}\equiv \frac{1}{\cO_i^0}
\bigg[\frac{d\cO_i}{d\ln\gamma_\alpha}\bigg]_{\gamma=\gamma^0}\!\!
\eea
$\tilde M_{ij}$ is a $s\times s$  matrix,  $\cJ_{i\alpha}$ is a $n\times s$ matrix and
\bea\label{LO}
L(\gamma)\equiv 
(2\pi)^{-\frac{s}{2}}\,(\det \tilde M)^{-\frac{1}{2}}
\,\exp\big[-1/2\,\,\tilde\gamma_\alpha\, \tilde M^{-1}_{\alpha\beta}
\,\tilde\gamma_\beta\big],
\qquad
\tilde\gamma_\alpha\equiv\gamma_\alpha/\gamma_\alpha^0-1.
\eea

\medskip\noindent
$L(\gamma)$ is the   normalized  distribution of  $\gamma_\alpha$ about 
 central $\gamma_\alpha^0$ that maximize it and contains correlations.
Eq.(\ref{log}) has  similarities  to eq.(\ref{LW}), (\ref{zz}).

Let us examine the matrix $\tilde M$  and assume for simplicity that
$M_{ij}$ is diagonal, then
\medskip
\bea\label{tttt}
 \tilde M^{-1}_{\alpha\beta}
\!=\!
 \sum_{i=1}^n\Big\{\Big(\frac{d (\cO_i/\sigma_i)}{d \ln\gamma_\alpha}
 \Big)
 \Big(\frac{d (\cO_i/\sigma_i)}{d \ln\gamma_\beta}
 \Big)\Big\}_{\gamma=\gamma^0},\qquad \alpha, \beta=1,2,....s.
\eea

\medskip\noindent
This expression   shows the relevance  of
 the variations   of $\cO_i$ ``normalized''  to their $\sigma_i$,
 which is somewhat expected on physical grounds.
Further,  $\cO_i$ are functions of $v(\gamma)$, 
$\cO_i\!=\!\cO_i(\gamma, v(\gamma))$, which
is relevant in establishing the relation of this matrix to the 
traditional fine-tuning. As a result of this dependence,
the matrix $\tilde M^{-1}$  contains {\it new terms} 
 \medskip
\bea\label{ww}
\tilde M^{-1}_{\alpha\beta}=\tilde M^{-1}_{\alpha\beta}\Big\vert_{v=const}&+&
\sum_{i=1}^s\! \Big\{\Big(\frac{\partial\cO_i/\sigma_i}{\partial\ln v}\Big)^2\!
\Big(\frac{\partial \ln v}{\partial \ln \gamma_\alpha}\Big)
\Big(\frac{\partial \ln v}{\partial \ln \gamma_\beta}\Big)\Big\}_{\gamma=\gamma^0}
\nonumber\\[4pt]
&+&
\sum_{i=1}^s\! \Big\{\Big(\frac{\partial\cO_i/\sigma_i}{\partial\ln v}\Big)\!
\Big(\frac{\partial \ln v}{\partial \ln \gamma_\alpha}\Big)
\Big(\frac{\partial \cO_i/\sigma_i}{\partial \ln \gamma_\beta}\Big)+(\alpha\leftrightarrow \beta)
\Big\}_{\gamma=\gamma^0},
\eea

\medskip\noindent
which are not  present in the traditional approach of  numerical data fits 
in which $v$ is actually a  {\it constant} (fixed input)
\cite{Bechtle}. 
So each entry  $\tilde M^{-1}_{\alpha\beta}$ 
{\it automatically} contains the  EW fine-tuning represented by the partial derivatives
of $v$ with respect to
 $\gamma_{\alpha},\gamma_\beta$, from all observables that depend on $v$!

This is an interesting result and suggests it is worth studying
other properties of $\tilde M$.
First, the trace 
\medskip
\bea
\Tr \,\tilde M^{-1}=
\sum_{i=1}^n
\sum_{\alpha=1}^s
\Big(\frac{d \cO_i/\sigma_i}{d \ln\gamma_\alpha}\Big)^2_{\gamma=\gamma^0}
= \sum_{i=1}^n \Big(\frac{\partial \cO_i/\sigma_i}{\partial\ln v}\Big)^2_{\gamma=\gamma^0}
\times\underbrace{\sum_{\alpha=1}^s 
\Big(\frac{\partial\ln v}{\partial\ln\gamma_\alpha}\Big)^2_{\gamma=\gamma^0}}_{\Delta^2}
+\cdots,
\eea

\medskip\noindent
contains terms  proportional to the traditional EW fine-tuning (second sum above),
with contributions from all $\cO_i$ that depend on $v(\gamma)$. 
This correction is also missed if $v$ is a constant
while retaining only the  explicit dependence of $\cO_i$ on $\gamma$. 
These results can also be examined for the single observable case, 
such as $m_Z$, $m_h$, $m_H$. 
Being invariant under the choice of basis (of parameters) 
the Trace has  some physical meaning and then so does the EW 
fine-tuning that emerges from it. 
So  $\tilde M$ with $v=v(\gamma)$ seems  more  fundamental than
the fine-tuning that was introduced in the past on physical grounds. 
These observations are easily  extended if the initial $M_{ij}$ is not diagonal.

The conclusion is that the ``usual'' EW fine-tuning  is automatically present
 in the analysis of precision data fits provided that one includes the EW scale
as $v=v(\gamma)$ predicted by SUSY. This  result has not
 been investigated numerically\footnote{It may be possible that
numerical studies account for this effect in a different way. With 
$v$ ($m_Z^0$) a fixed input, the EW minimum condition brings instead a dependence
say $\mu=\mu(\gamma_\alpha)$
where $\gamma_\alpha$ denote parameters other than $\mu$. With this 
dependence it is possible to account for the above effect, but 
the presence in the covariance matrix  of the EW fine-tuning  seen above
 is not manifest and is overlooked.};  one can re-evaluate
the precision data fits  to treat $v$ as  a function $v=v(\gamma)$ and 
include observables that depend on it ($m_Z$, $m_h$,  etc)  
together with the additional likelihood ``cost'' it could bring. 
In this picture, the traditional EW fine-tuning per 
se and its numerical  value  may be  less relevant since  $\tilde M$ 
contains the  information related to these.

Further, one can also consider the determinant of the (inverse) 
covariance matrix $\tilde M^{-1}$ in the basis of the fundamental parameters. 
It is actually more relevant to consider this determinant
relative to that of the initial matrix $M$, 
and  this gives exactly $\Delta$ of eq.(\ref{pp}). This factor related
 the normalised $L(\cO\vert\gamma)$ and $L(\gamma)$ in
 eq.(\ref{log}) and it is a measure of their relative width\footnote{
In information theory \cite{it} $\ln\Delta$ is interpreted as the change of the 
differential entropy when going from a multivariate Gaussian 
distribution (of observables $\cO_i$) to another one (here of parameters $\gamma_\alpha$).}.  
For simplicity, take the case when 
the number of observables $\cO_i$ equals that of the parameters
$\gamma_\alpha$  ($n\!=\!s$). Then
\medskip
\bea
\Delta\!=\!(\det \cJ^T\,\cJ)^{1/2}
\qquad
 \eea

\medskip\noindent
In particular, for two observables (say $m_h$ and $m_Z$) and two parameters:
\medskip
\bea\label{ind}
\Delta=
\Delta_1\,\Delta_2\,
\big[1-\xi_{12}\big]^{1/2},
\eea
where
\medskip
\bea\label{x}
\xi_{12}\equiv
\frac{1}{\Delta_1^2\,\Delta_2^2}\,\big[ \cJ_{1\alpha}\,\cJ_{2\alpha}\big]^2
\qquad
\Delta_k
 =\Big\{\sum_\alpha \Big(\frac{\cO_k(\gamma^0)}{\cO_k^{0}}
 \frac{d \ln \cO_k}{d\ln \gamma_\alpha}
\Big)^2_{\gamma=\gamma^0}\Big\}^{1/2}, \,\,\,k=1,2.
\eea

\medskip
\noindent
With $\cO_i$ functions of $v(\gamma)$, the individual fine-tunings $\Delta_{1,2}$ of 
$\cO_{1,2}$ include the EW fine-tuning and are part of this more general
$\Delta$. So we see again that fine tuning of the observables and of the EW scale
 is included by  the covariance matrix\footnote{
$\Delta$ of eq.(\ref{ind}) is smaller then
when the variations of $\cO_i$ are 
orthogonal ($\xi_{12}=0$ i.e. independent $\cO_i$). }.
The above results answered question $Q_2$ in Introduction.

From our result above it is clear that 
the usual criterion of a good fit in a model
 $\chi^2/n_{df}\approx 1$ ($\chi^2\equiv-2\ln L$) 
imposed on this matrix in a numerical analysis of the EW data with $v=v(\gamma)$
should then automatically  take fine tuning  into account; this can then 
bring bounds on EW fine-tuning.  
In a simplified set-up and under additional assumptions, this procedure  was 
used  in  \cite{Ghilencea:2013hpa,Ghilencea:2013fka} to set bounds on   $\Delta$.
We do not pursue this method here.

In the following let us be more general and analyse instead
 the s-standard deviation confidence interval, 
defined by the surface\footnote{The value of
$s$ depends on the number of degrees of freedom $n_{df}$.}:
\bea\label{op1}
-2\ln L(\gamma^\prime)\leq  -2\ln L(\gamma^0)+s^2,
\eea
with  $L(\gamma^0)=L_{max}$  and $\gamma^\prime=\gamma^\prime(s)$. In $\chi^2$-language this
becomes $\delta\chi^2=\chi^2-\chi^2_{min}\leq s^2$. In our approximation 
this condition becomes, from eqs.(\ref{LO}), (\ref{tttt}) 
\medskip
\bea\label{ol}
\delta\chi^2=\tilde\gamma_\alpha \,\tilde M^{-1}_{\alpha\beta}\, \tilde\gamma_\beta
=
\sum_{i=1}^n \Big\{\Big(\frac{d \cO_i/\sigma_i}{d\ln \gamma_\alpha}\Big)_{\gamma=\gamma^0} 
(\gamma_\alpha^\prime/\gamma_\alpha^0-1)\Big\}^2
\leq s^2
\eea

\medskip\noindent
(implicit sum over $\alpha$). 
With $\cO_i$  a function of $v(\gamma)$, for  fixed  $s$ 
and assuming that $\vert\gamma_\alpha^\prime-\gamma_\alpha^0\vert\geq \sigma_{th,\alpha}>0$
this condition brings a bound on the EW fine-tuning.
We introduced `ad-hoc'
$\sigma_{th,\alpha}$ as a theoretical error of computing  $\gamma_\alpha^0$, 
from the maximal likelihood (min $\chi^2$) condition
in which theoretical values of observables are affected by ignored 
higher loops errors\footnote{An example
is that of 2-3 GeV higher loop error mentioned in Section 1.2 for $m_h$.}
(including effects of the   RG flow for  $\gamma_\alpha$  
 these observables depend on).

From  inequality  (\ref{ol}) for one observable only ($m_Z$):
\medskip
\bea\label{op2}
\Delta\leq \frac{s\,\sigma_z}{m_Z(\gamma^0)}\Big\vert \frac{n^{\alpha}\,(\gamma_\alpha^\prime-
\gamma_\alpha^0}{\gamma_\alpha^0}\Big\vert^{-1}
\eea
with $\Delta$ as in eq.(\ref{Delta}). This also gives
\bea\label{pp3}
\Delta \leq \frac{s\,\sigma_z}{m_Z(\gamma^0)}\Big\{\sum_{\alpha\geq 1}^s
\Big(\frac{\gamma^0_{\alpha}}{\sigma_{th,\alpha}}\Big)^2\Big\}^{1/2}
 \eea
This bound  is similar to that discussed  in Section~\ref{qqq}
 and  eq.(\ref{sigma}) and depends on the
 experimental and theoretical errors. The strength of this bound 
depends on the values of  $\sigma_{th,\alpha}$ and  $\sigma_z$ (more generally $\sigma_i$ of all
 $\cO_i$) and can be  enhanced by the presence of more observables, see eq.(\ref{ol}).
Finally, using this approach in precision data fits of the EW data,  a plot of the lhs of eq.(\ref{ol}) 
giving $\delta\chi^2$, as a function of $\Delta$, for current value of the higgs mass could
 illustrate  the role that fine-tuning plays in deciding if a model
is realistic. Bounds (\ref{op2}), (\ref{pp3}) can be generalised when original 
$M_{ij}$ is not diagonal ({\it i.e.} when eq.(\ref{tttt}) is not valid anymore).

The  matrix $\tilde M$ has another interesting feature.
A large traditional EW fine-tuning, which is more a problem
of supersymmetry breaking than of supersymmetry itself, can signal
 that our theory of soft terms ($\gamma_\alpha$) is incomplete. 
As mentioned, relations among soft masses
(such as GUT relations among the  gaugino masses, etc) can reduce its value.  
There is then the possibility that
 some parameters $\gamma_\alpha$ could be related. 
Such relations can be captured by the matrix $\tilde M$ as off-diagonal 
entries.  This means that  properties of this matrix  can help 
us identify  the   fundamental $\gamma_\alpha$
under the constraints of the model.
Indeed, there exists the so-called
{\it global} correlation coefficient of one such  parameter ($\gamma_\alpha$) with 
the rest,  defined as 
\medskip
\bea
\rho_\alpha=\sqrt{1-\big[ \tilde M_{\alpha\alpha}\,(\tilde M^{-1})_{\alpha\alpha}\big]^{-1}},\qquad
0\leq \rho_\alpha\leq 1
\eea

\medskip\noindent
$\rho_\alpha$ measures the total amount of 
correlation between  $\gamma_\alpha$ and  all other parameters 
$\gamma_\beta$ ($\beta\not=\alpha$).  If $\rho_\alpha=0$ then $\gamma_\alpha$ 
is an independent  variable  while if $\rho_\alpha\rightarrow 1$ there is full
correlation  of $\gamma_\alpha$ with  one linear combination  of the other 
parameters; this is captured by the off-diagonal terms when 
inverting the matrix.  $\rho_\alpha$ could help  a 
better understating of the SUSY breaking soft terms.
In this sense $\rho_\alpha$, $\alpha=1,2..s$ could also  be used  as a 
new measure of EW fine-tuning defined as  $\tilde \Delta=\max\vert\rho_\alpha\vert$.

Interestingly, another coefficient was discussed previously, for the correlation  
between  pairs of parameters, $\rho_{\alpha\beta}$ \cite{Bechtle}; this was used
to define a new measure of EW fine-tuning as 
$\max\vert\rho_{\alpha\beta}\vert$, where
$\rho_{\alpha\beta}\sim \tilde M_{\alpha\beta}/(\sigma_\alpha\sigma_\beta)$.
This  measure was reached on physical grounds and  again supports 
the connection of  the matrix $\tilde M$ to  fine-tuning, emphasized here.  
We insist  that  when one computes the coefficient  $\rho_\alpha$
as well as $\rho_{\alpha\beta}$ the EW scale $v$ be regarded as a function 
of $\gamma$'s, to reflect this original prediction of SUSY.

To conclude, a traditional frequentist analysis of the EW observables 
(including $m_Z$) with the constraint that  $v$ is a function  $v=v(\gamma)$ 
is a test  that remains true to the original 
motivation of SUSY. The EW fine-tuning due to all observables is  
automatically captured by the covariance matrix
if $v=v(\gamma)$ and in this case there may be no need to  discuss $\Delta$ separately.
Upper bounds on the EW fine-tuning emerge, as shown in eq.(\ref{op2}) from the:

\noindent
- standard deviation interval constraint discussed 
(imposed on this matrix), see eq.(\ref{op1})

\noindent
-  the ignored higher-loops error ($\sigma_{th,\alpha}$)
affecting the theoretical calculation of the UV parameters $\gamma_\alpha$ of the Lagrangian\footnote{
Upper bounds on $\Delta$ also emerge from the criterion  $\chi^2/n_{df}\approx 1$ (good fit) 
 \cite{Ghilencea:2013hpa,Ghilencea:2013fka}, not discussed here.}.

This discussion  relied on a Taylor expansion of $\cO_i$ to linear order
which, although illustrative for our purpose, is not acceptable for precision studies. 
A numerical approach, with $v=v(\gamma)$, can avoid this approximation.

\section{Conclusions.}

Unlike the SM, its supersymmetric versions
stabilize the EW scale  $v\sim m_Z$ at the quantum level and
 {\it predict} that   $v$ is a   derived quantity, function of 
the SUSY UV parameters  $\gamma_\alpha$ (soft masses, couplings and $\mu$), 
so $v\!=\!v(\gamma)$.
Whether this SUSY prediction successfully recovers 
its experimental value is the natural test of  this theory. 
This view remains true to the original motivation of SUSY.
Past  estimates showed that  fixing the EW 
scale to its measured value affects the  likelihood to fit the data 
by a factor related to  the EW fine-tuning. 
Here we examined this problem in a different, more general approach.

The  result is that the covariance matrix in the basis of the
parameters $\gamma_\alpha$  automatically encodes information about the 
 EW fine-tuning {\it  provided that} the EW scale is regarded as a function 
$v\! = \!v(\gamma)$ (rather than a constant).
Note that such connection between this  matrix and the 
EW fine tuning was not previously
examined in the literature, even though each of these aspects were studied separately.
Further, the Trace  of the inverse of the covariance matrix  and its determinant
also contain the EW fine tuning due to {\it all} EW observables that depend on $v$.
This indicates that the EW fine-tuning is somewhat less fundamental since 
this matrix includes its effects through the variations of the EW scale $v$
with respect to $\gamma_\alpha$, (closely related to $\Delta$).

As a result, the  evaluation of the traditional EW fine-tuning per se 
is then less relevant for the viability of a model  {\it as long as  with  
$v=v(\gamma)$} a good $\delta\chi^2$ of the  observables (including $m_Z$)  is still 
possible in that model, within the theoretical approximation (loop order) 
considered. From this condition and approximation one can subsequently infer 
numerical bounds on the EW  fine tuning.
More explicitly, the deviation $\delta\chi^2$ from the minimal value $\chi^2_{min}$ 
is  affected by the EW fine-tuning; so for a s-standard deviation confidence interval (region)
and a given theoretical error (loop order), a bound on the traditional measure 
of the EW fine tuning (``in quadrature'') was obtained (eq.(\ref{op2}), (\ref{pp3})).
 This eliminates subjective
criteria for  ``acceptable'' numerical values for $\Delta$.

 At present the above effect seems to be 
overlooked in the precision data fits in the frequentist approach where
$v$  is actually a fixed, input constant so the ``fine-tuning''-related 
 corrections to the covariance matrix shown in 
the text (eq.(\ref{ww}) seem to be ignored; or they  are indeed included 
but in a way which does not make manifest the role of the EW fine tuning that we showed. 
This effect  needs further 
numerical investigation. Our result also answers how to compare 
a model with a good fit of the data but significant EW fine tuning against 
a model with nearly-as-good a fit but less
EW fine tuning: since fine-tuning effects are included in the covariance matrix,
one simply chooses the model with the best fit 
obtained {\it with} $v=v(\gamma)$.
This answers our remaining question ($Q_3$) in Introduction.

A large EW fine-tuning can  be an indication of
our ignorance of the details of the SUSY breaking mechanism and 
of the lack of  a theory of soft terms. It is known that 
symmetries that relate the soft terms can  reduce its value.
The  global correlation coefficient of the covariance matrix can show
if a particular parameter $\gamma_\alpha$ is correlated with 
a combination of the rest.
This could help one  trace the more fundamental SUSY parameters
 and better understand the relation of fine-tuning to supersymmetry breaking.

\bigskip
\bigskip
\bigskip
\bigskip

\noindent
{\bf Acknowledgements: }
The author thanks Stefan Pokorski for many interesting discussions.
He also thanks Philip Bechtle, Werner Porod and Tim Stefaniak for a related
discussion. This work was supported by 
a grant of the Romanian National Authority for Scientific Research, CNCS - UEFISCDI, 
project number PN-II-ID-PCE-2011-3-0607 and in part by Programme 
`Nucleu'  PN 09 37 01 02.

\end{document}